\documentclass[parskip=half]{scrartcl}

\pdfoutput=1

\usepackage{amssymb}
\usepackage{amsmath}
\usepackage{enumitem}
\usepackage{xcolor}
\definecolor{Red}{rgb}{0.5,0,0}
\definecolor{Blue}{rgb}{0,0,0.5}
\definecolor{Black}{rgb}{0,0,0}
\usepackage[numbers,round]{natbib}
\usepackage[nogin]{Sweave}
\usepackage{subfig}
\usepackage{hyperref}
\hypersetup{%
  hyperindex = {true},
  colorlinks = {true},
  linktocpage = {true},
  plainpages = {false},
  linkcolor = {Blue}, % {Black},
  citecolor = {Blue}, % {Black},
  urlcolor = {Red},  % {Black},
}

\setdescription{font=\normalfont}

\newcommand{\dsvar}[1]{\textit{#1}}

\setcapindent{0em}

\begin{document}

\title{Archetypal Athletes}
\author{Manuel J. A. Eugster\\
  {\small Ludwig-Maximilans-Universt{\"a}t M{\"u}nchen}}
\date{}

\maketitle

\begin{abstract}
  \noindent Discussions on outstanding---positively and/or
  negatively---athletes are common practice. The rapidly grown amount of
  collected sports data now allow to support such discussions with state
  of the art statistical methodology. Given a (multivariate) data set
  with collected data of athletes within a specific sport, outstanding
  athletes are values on the data set boundary. In the present paper we
  propose archetypal analysis to compute these extreme values. The
  so-called archetypes, i.e., archetypal athletes, approximate the
  observations as convex combinations. We interpret the archetypal
  athletes and their characteristics, and, furthermore, the composition
  of all athletes based on the archetypal athletes. The application of
  archetypal analysis is demonstrated on basketball statistics and
  soccer skill ratings.

  \bigskip

\end{abstract}

% \input{content}
%%%%%%%%%%%%%%%%%%%%%%%%%%%%%%%%%%%%%%%%%%%%%%%%%%%%%%%%%%%%%%%%%%%%%%
\section{Introduction\label{intro}}

``Dirk Nowitzki is the best basketball player. No, it's Kevin
Durant!''. ``Christiano Ronaldo is the number one, Lionel Messi number
two soccer player in the world''. ``Ronaldinho is the better dribbler,
but Zinédine Zidane is faster''. These and similar statements can be
found in almost all discussions on sports and the practicing
athletes. They are interesting to debate, but they are also having a
great impact on many (managerial) decisions---from a coach's tactical
specification via engagements of new players through to a company's
selection of a brand ambassador. The consequence is the collection of
more and more sports data. A large number of statistics (the
variables) per sports and athletes (the observations) are measured to
investigate such statements using state of the art statistical
methodology.

The foundations of statements like the introductory examples are
constructed orders of the athletes (maybe implicit). Given that no
uniquely defined strict order (and therefore no minimum and maximum)
exists for observations with more than one dimension, most approaches
are based on an appropriate reduction of the collected statistics to
the one-dimensional space (where a strict order exists). General
methods are for example ordination methods like multidimensional
scaling and principal components analysis
\citep[e.g.,][]{Hastie+Tibshirani+Friedman@2009}; specialized methods
are for example the EA Sports Player Performance Index (previously
Actim index) for soccer \citep{McHale+Scarf@2005} and the Total Player
Rating for baseball \citep{Thorn+Palmer@1984}. Obviously, the
reduction to the one dimensional space implies the loss of
information---it enables a simply ranking of the athletes, but in case
of an objective evaluation it might cause discrepancies.

% Simple example:
% <<>>=
% d <- matrix(c(c(1, 1), c(5, 5), c(2, 3), c(3, 2)), ncol = 2, byrow = TRUE)
% plot(d, col = 1:4, pch = 19)
% s <- cmdscale(dist(d), k = 1)
% plot(s, rep(1, 4), col = 1:4, pch = 19)
% @

Archetypal analysis has the aim to find a few, not necessarily
observed, extremal observations (the archetypes) in a multivariate
data set such that all the data can be well represented as convex
combinations of the archetypes. The archetypes themselves are
restricted to being convex combinations of the individuals in the data
set and lie on the data set boundary, i.e., the convex hull. This
statistical method was first introduced by \citet{Cutler+Breiman@1994}
and has found applications in different areas, e.g., in economics
\citep{Li+Wang+Louviere+Carson@2003, Porzio+Ragozini+Vistocco@2008},
astrophysics \citep{Chan+Mitchell+Cram@2003} and pattern recognition
\citep{Bauckhage+Thurau@2009}.

Archetypes can be seen as data-driven extreme values. In sports data,
these extreme values are the archetypal athletes; athletes which are
outstanding---positively and/or negatively---in one or more of the
collected statistics. For interpretation, we identify the archetypal
athletes as different types of ``good'' and ``bad'', and set the
observations in relation to them. Statements like ``Dirk Nowitzki is
the best basketball player'' are then easily verified---the athlete
has to be an archetypal athlete (or its nearest
observation). Furthermore, statements like ``Ronaldinho is the better
dribbler'' are verified by not only interpreting the observations'
nearest archetypes but their (convex) combinations of all archetypes.

The paper is organized as follows. In Section~\ref{archetypes} we
outline archetypal analysis by introducing the formal optimization
problem. We illustrate the idea of archetypal analysis using a
two-dimensional subset of NBA player statistics from the season
2009/2010. In Section~\ref{athletes} we then identify and discuss
archetypal athletes for two popular sports. Section~\ref{basketball}
extends the illustrative NBA example and computes archetypal
basketball players using common statistics from the season
2009/2010. Section~\ref{soccer} computes archetypal soccer players of
the German Bundesliga, the English Premier League, the Italian Lega
Serie A, and the Spanish La Liga using skill ratings (at the time of
September 2011). Finally, in Section~\ref{conclusion} the conclusions
are given. All data sets and source codes for replicating our analyses
are freely available (section on computational details on
page~\pageref{compdetails}).

%%%%%%%%%%%%%%%%%%%%%%%%%%%%%%%%%%%%%%%%%%%%%%%%%%%%%%%%%%%%%%%%%%%%%%
\section{Archetypal analysis\label{archetypes}}

Consider an $n \times m$ matrix $X$ representing a multivariate data
set with $n$ observations and $m$ attributes. For given $k$ the
archetypal problem is to find the matrix $Z$ of $k$ $m$-dimensional
archetypes. More precisely, to find the two $n \times k$ coefficient
matrices $\alpha$ and $\beta$ which minimize the residual sum of
squares
\begin{gather}
  \label{formula:atypes}
  \text{RSS} = \|X - \alpha Z^{\top}\|_2  \text{ with } Z = X^{\top}
  \beta
\end{gather}
subject to the constraints
\begin{gather*}
  \sum_{j=1}^{k} \alpha_{ij} = 1 \text{ with } \alpha_{ij} \geq 0
  \text{ and } i = 1, \ldots, n\text{,}\\
  \sum_{i=1}^{n} \beta_{ji} = 1 \text{ with } \beta_{ji} \geq 0 \text{
    and } j = 1, \ldots, k\text{.}
\end{gather*}
The constraints imply that (1)~the approximated data are convex
combinations of the archetypes, i.e., $X = \alpha Z^\top$, and (2)~the
archetypes are convex combinations of the data points, i.e., $Z =
X^\top \beta$. $\|\cdot\|_2$ denotes the Euclidean matrix norm.

\citet{Cutler+Breiman@1994} present an alternating constrained least
squares algorithm to solve the problem: it alternates between finding
the best $\alpha$ for given archetypes $Z$ and finding the best
archetypes $Z$ for given $\alpha$; at each step several convex least
squares problems are solved, the overall $\mbox{RSS}$ is reduced
successively. Through the definition of the problem, archetypes lie on
the boundary of the convex hull of the data. Let $N$ be the number of
data points which define the boundary of the convex hull, then
\citet{Cutler+Breiman@1994} showed: if $1 < k < N$, there are $k$
archetypes on the boundary which minimize $\mbox{RSS}$; if $k = N$,
exactly the data points which define the convex hull are the
archetypes with $\mbox{RSS} = 0$; and if $k = 1$, the sample mean
minimizes the $\mbox{RSS}$. In practice, however, these theoretical
results can not always be achieved \citep{Eugster+Leisch@2009}.
Furthermore, there is no rule for the correct number of archetypes $k$
for a given problem instance. A simple method to determine the value
of $k$ is to run the algorithm for increasing numbers of $k$ and use
the ``elbow criterion'' on the $\mbox{RSS}$ where a ``flattening'' of
the curve indicates the correct value of $k$. For detailed
explanations we refer to \citet[][on the original
algorithm]{Cutler+Breiman@1994}, \citet[][on numerical issues,
stability, and computational complexity]{Eugster+Leisch@2009}, and
\citet[][on robustness]{Eugster+Leisch@2011}.

\begin{figure}[t]
  \centering
  \subfloat[][]{\label{fig:nba2d-data}
\includegraphics{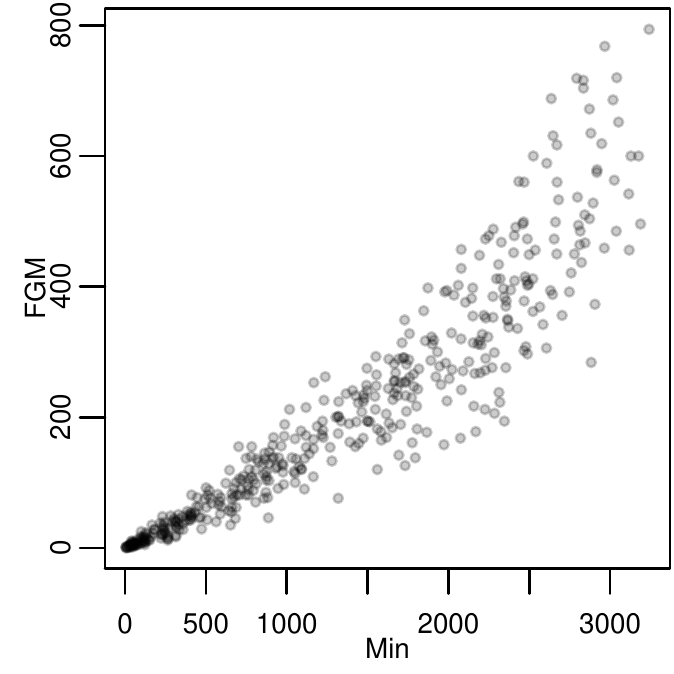}
  }
  \subfloat[][]{\label{fig:nba2d-a3}
\includegraphics{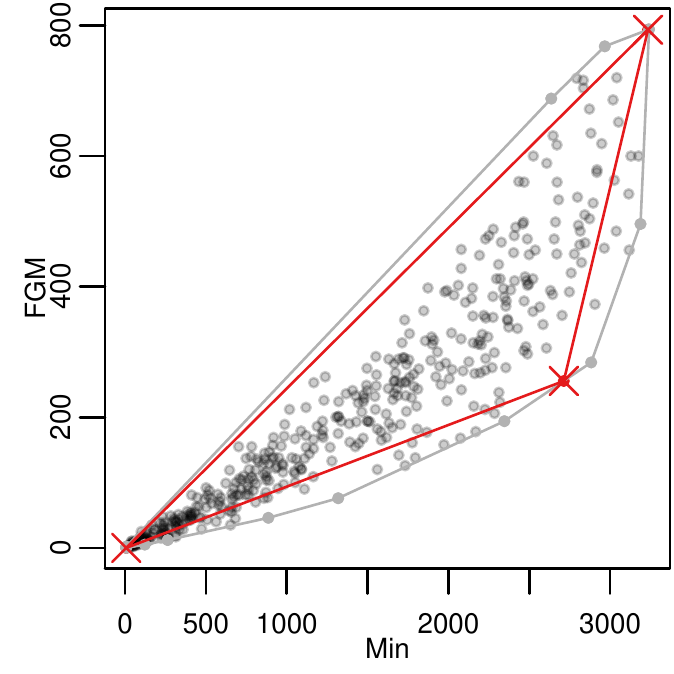}
  }
  \caption{(a)~Data set of two NBA player statistics from the season
    2009/2010. (b)~Convex hull (gray) and the corresponding three
    archetypes solution (red).}
  \label{fig:nba2d}
\end{figure}

In order to illustrate archetypal analysis, we use a two-dimensional
subset of the NBA player statistics from the season 2009/2010 which we
analyze in Section~\ref{basketball}: the two variables are
\dsvar{total minutes played} (\dsvar{Min}) and \dsvar{field goals
  made} (\dsvar{FGM}) of 441 players, i.e., we investigate ``the
score efficiency''. Figure~\ref{fig:nba2d-data} shows the data
set. The majority of players are in the range $[0, 1000]$ of
\dsvar{Min} and $[0, 200]$ of \dsvar{FGM}. With increasing
\dsvar{Min}, the variance in \dsvar{FGM} increases and the shape of
the data set suggests the estimation of three
archetypes. Figure~\ref{fig:nba2d-a3} visualizes the $k = 3$
archetypes solution (red), together with the data's convex hull
(gray). We see that this archetypes solution is a reasonable
approximation of the convex hull (note that the archetypes do not have
to be observed data points). Using this solution, the data points
inside the archetypes solution are exactly approximated, the data
points outside the archetypes solution are approximated with an error,
as they are projected on the hull of the archetypes solution.

The three archetypes can be interpreted as follows. Concerning these
two variables \dsvar{total minutes played} (\dsvar{MIN}) and
\dsvar{field goals made} (\dsvar{FGM}), three types of extreme players
are in the data set:
\begin{description}
  \item [Archetype~1] is the natural ``maximum'' with high values in
    all variables ($\mbox{\dsvar{Min}} = 3234$,
    $\mbox{\dsvar{FGM}} = 793$); this archetype
    represents a type of ``\textit{good}'' scorer.

  \item [Archetype~2] is the natural ``minimum'' with low values in
    all variables ($\mbox{\dsvar{Min}} = 7$,
    $\mbox{\dsvar{FGM}} = 0$); this archetype
    represents a type of ``\textit{bad}'' scorer.

  \item [Archetype~3] is another extreme value with a high number of
    \dsvar{Min} but a (relatively) low number of \dsvar{FGM}
    ($\mbox{\dsvar{Min}} = 2713$, $\mbox{\dsvar{FGM}} =
    256$); this archetype represents another type of
    ``\textit{bad}'' scorer (i.e., an ineffective scorer).
\end{description}
Note that there is no archetype with a low number of \dsvar{Min} and
a high number of \dsvar{FGM}; such an archetype would represent
another type of ``\textit{good}'' scorer (i.e., an effective
scorer). An important aspect of the interpretation is, that it is
conditioned on the given data; e.g., the number \dsvar{field goals
  made} obviously is related to the position and tactical orientation
of a player, but these information are not available in this
illustrative data set and therefore cannot contribute to the
interpretation of ``\textit{good}'' and ``\textit{bad}'' players.

\begin{figure}[t]
  \centering
  \subfloat[][]{
\includegraphics{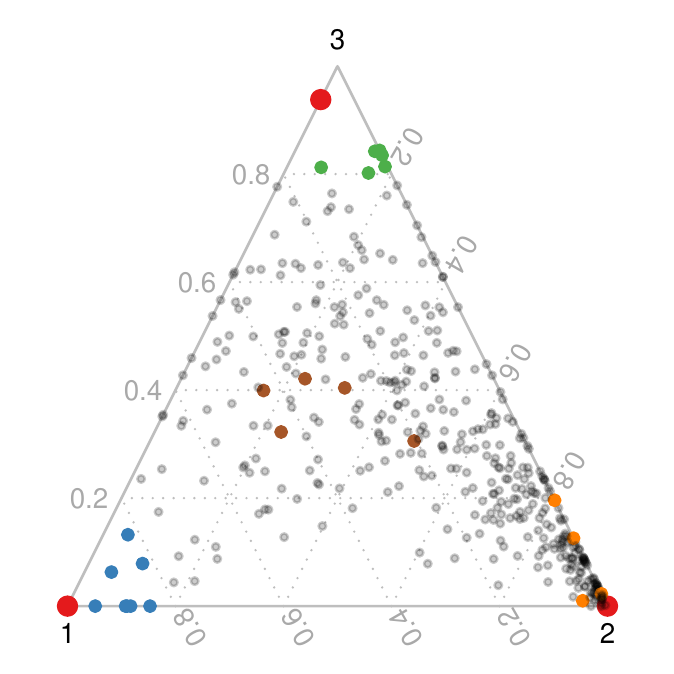}
  }
  \subfloat[][]{
\includegraphics{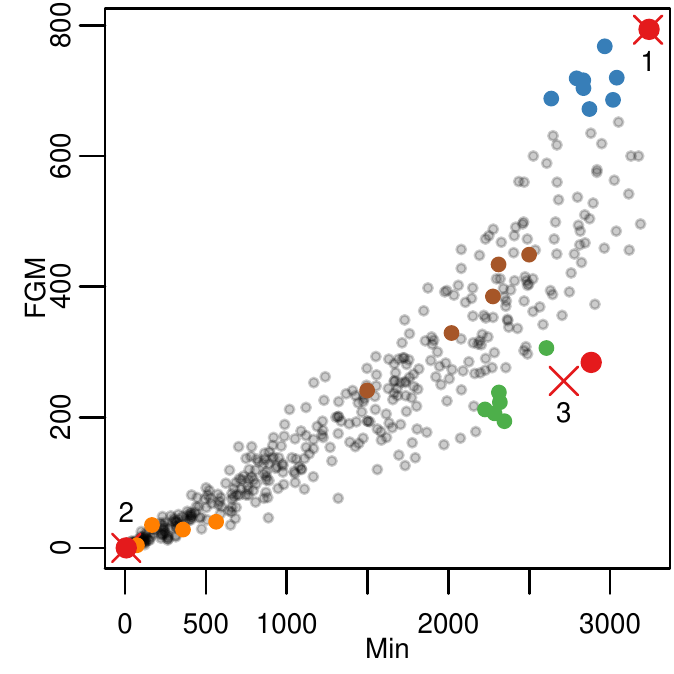}
  }
  \caption{(a)~Visualization of the $\alpha$ coefficients using a
    ternary plot and (b)~the data set in case of the $k = 3$
    archetypes solution. The red dots are the archetypes' nearest
    players; dots colored with blue, orange, and green are players
    where Archetype~1, 2, and 3 contribute more than $0.8$.}
  \label{fig:nba2d-a3-alphas}
\end{figure}

Having identified the possible extreme values within the given data
set, the next step is to set the observations in relation to them. The
$\alpha$ coefficients of the archetypal problem
(Formula~\ref{formula:atypes}) define how much each archetype
contributes to the approximation of each individual observation (as
convex combination). This allows the assignment of the observations to
their nearest archetypes and, consequently, the identification of the
most archetypal observation(s). Figure~\ref{fig:nba2d-a3-alphas} shows
the corresponding ternary plot of the $\alpha$ coefficients for the
above $k = 3$ archetypes solution. The three players (red points)
nearest to the respective archetypes (red crosses) are:
\begin{center}
% latex table generated in R 2.13.1 by xtable 1.5-6 package
% Tue Sep 13 10:27:53 2011
\begin{tabular}{rlllrrrrr}
  \hline
 & Name & Team & Role & Min & FGM & $\alpha_{\cdot1}$ & $\alpha_{\cdot2}$ & $\alpha_{\cdot3}$ \\
  \hline
Archetype 1 & Kevin Durant & OKL & SF & 3241 & 794 & 1.00 & 0.00 & 0.00 \\
  Archetype 2 & Dwayne Jones & PHO & C &   7 &   0 & 0.00 & 1.00 & 0.00 \\
  Archetype 3 & Jason Kidd & DAL & PG & 2883 & 284 & 0.06 & 0.00 & 0.94 \\
   \hline
\end{tabular}\end{center}
Archetype~1 and 3 have well-defined nearest observations; Archetype~2,
on the contrary, has a set of nearest observations and the concrete
player identification should be considered as a ``random'' selection
from the set of similar players.

We have identified Archetype~1 as the ``\textit{good}'' archetype in
this data setting---on this account, Kevin Durant can be
considered as the best scorer. To find other good scorers, we look at
the observations where Archetype~1 contributes more than $0.8$ (blue
points):
\begin{center}
% latex table generated in R 2.13.1 by xtable 1.5-6 package
% Tue Sep 13 10:27:53 2011
\begin{tabular}{lllrrrrr}
  \hline
Name & Team & Role & Min & FGM & $\alpha_{\cdot1}$ & $\alpha_{\cdot2}$ & $\alpha_{\cdot3}$ \\
  \hline
Kevin Durant & OKL & SF & 3241 & 794 & 1.00 & 0.00 & 0.00 \\
  Lebron James & CLE & SF & 2967 & 768 & 0.95 & 0.05 & 0.00 \\
  Kobe Bryant & LAL & SG & 2834 & 716 & 0.89 & 0.11 & 0.00 \\
  Dwyane Wade & MIA & SG & 2793 & 719 & 0.89 & 0.11 & 0.00 \\
  Dirk Nowitzki & DAL & PF & 3041 & 720 & 0.89 & 0.05 & 0.06 \\
  Amare Stoudemire & PHO & PF & 2835 & 704 & 0.88 & 0.12 & 0.00 \\
  Carmelo Anthony & DEN & SF & 2636 & 688 & 0.85 & 0.15 & 0.00 \\
  David Lee & NYK & C & 3018 & 686 & 0.82 & 0.05 & 0.13 \\
  Derrick Rose & CHI & PG & 2872 & 672 & 0.82 & 0.10 & 0.08 \\
   \hline
\end{tabular}\end{center}
We equivalently proceed for the other two archetypes: Players where
Archetype~3 contributes more than $0.8$ are Jason Kidd, Thabo Sefolosha, Earl Watson, Anthony Parker, Derek Fisher, Ron Artest, Marcus Camby
(green points). Five randomly selected players where Archetype~2
contributes more than $0.8$ are Ryan Bowen, Sean Marks, Ian Mahinmi, Jamaal Magloire, Quinton Ross (orange
points).

Observations toward the center of the data set are not approximated
by one archetype alone, but each archetype contributes a significant
fraction. The following five players, for example, are randomly
selected from the data sets' center (brown points):
\begin{center}
% latex table generated in R 2.13.1 by xtable 1.5-6 package
% Tue Sep 13 10:27:53 2011
\begin{tabular}{lllrrrrr}
  \hline
Name & Team & Role & Min & FGM & $\alpha_{\cdot1}$ & $\alpha_{\cdot2}$ & $\alpha_{\cdot3}$ \\
  \hline
Vince Carter & ORL & SG & 2310 & 434 & 0.44 & 0.23 & 0.32 \\
  Anthony Morrow & GSW & SG & 2019 & 329 & 0.28 & 0.31 & 0.40 \\
  C.j. Miles & UTA & SF & 1497 & 241 & 0.21 & 0.49 & 0.31 \\
  Paul Millsap & UTA & PF & 2275 & 385 & 0.35 & 0.23 & 0.42 \\
  Rodney Stuckey & DET & PG & 2499 & 449 & 0.44 & 0.16 & 0.40 \\
   \hline
\end{tabular}\end{center}
As we can see, based on the $\alpha$ coefficients of the players no
assignments to one of the archetypes are possible.

Besides setting the observations in relation to their nearest
archetype using the observations' highest $\alpha$, the interpretation
of all $\alpha$s of an observation is of interest as well. Suppose
that, for example, the data set describes skill ratings of players,
then the $\alpha$s can be interpreted as the players' compositions of
skills; see Section~\ref{soccer} for such an application of archetypal
analysis.

%%%%%%%%%%%%%%%%%%%%%%%%%%%%%%%%%%%%%%%%%%%%%%%%%%%%%%%%%%%%%%%%%%%%%%
\clearpage
\section{Archetypal athletes\label{athletes}}

Archetypal analysis in general enables to compute data-driven extreme
values and the corresponding observations' (convex) combinations of
all archetypes. In case of sports data this allows to identify and
interpret archetypal athletes. Furthermore, all athletes are set in
relation to the archetypal athletes and then can be ``evaluated''
according to them.

In this section we determine archetypal athletes for two popular
sports and their representative leagues. Section~\ref{basketball}
extends the illustrative two-dimensional example and computes
archetypal basketball players with common statistics from the NBA
season 2009/2010. Section~\ref{soccer} determines archetypal soccer
players of the German Bundesliga, the English Premier League, the
Italian Lega Serie A, and the Spanish La Liga using skill ratings (at
the time of September 2011).

\subsection{Archetypal basketball players\label{basketball}}

\begin{figure}
  \centering
  \includegraphics{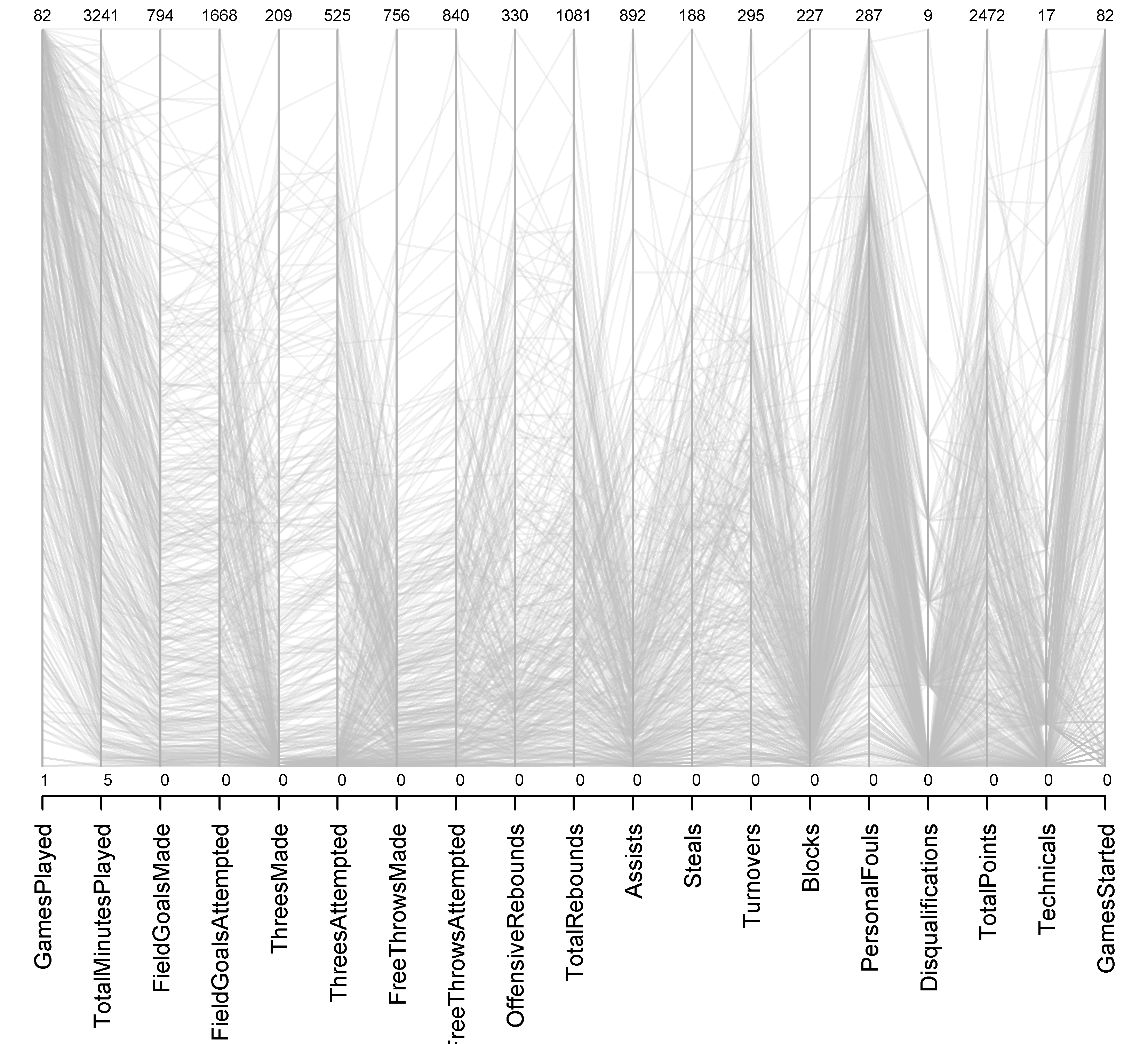}
  \caption{Parallel coordinates plot of the statistics of
    $441$ players from the NBA season 2009/2011.}
  \label{fig:nba-data}
\end{figure}

\begin{figure}
  \centering
\includegraphics{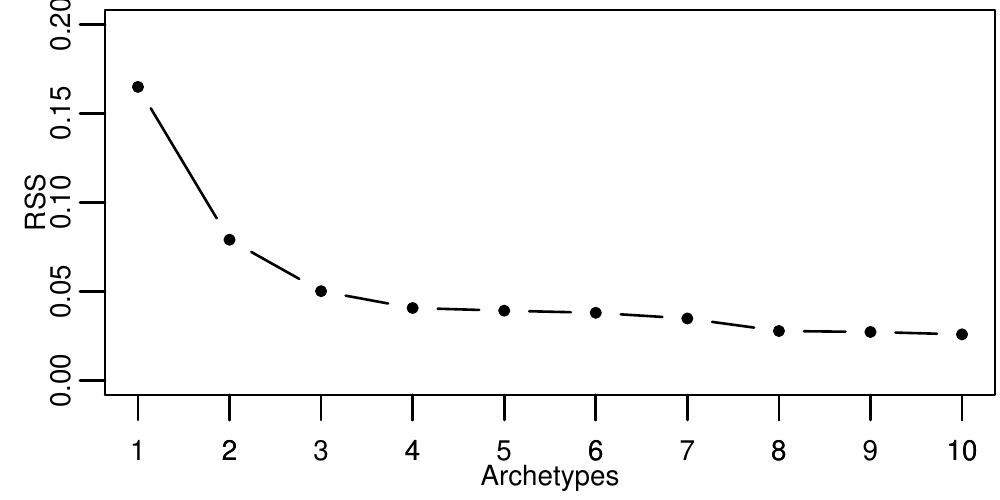}
  \caption{Scree plot of the residual sum of squares for $1$ to $10$
    archetypes.}
  \label{fig:nba-rss}
\end{figure}

\begin{figure}
  \centering
\includegraphics{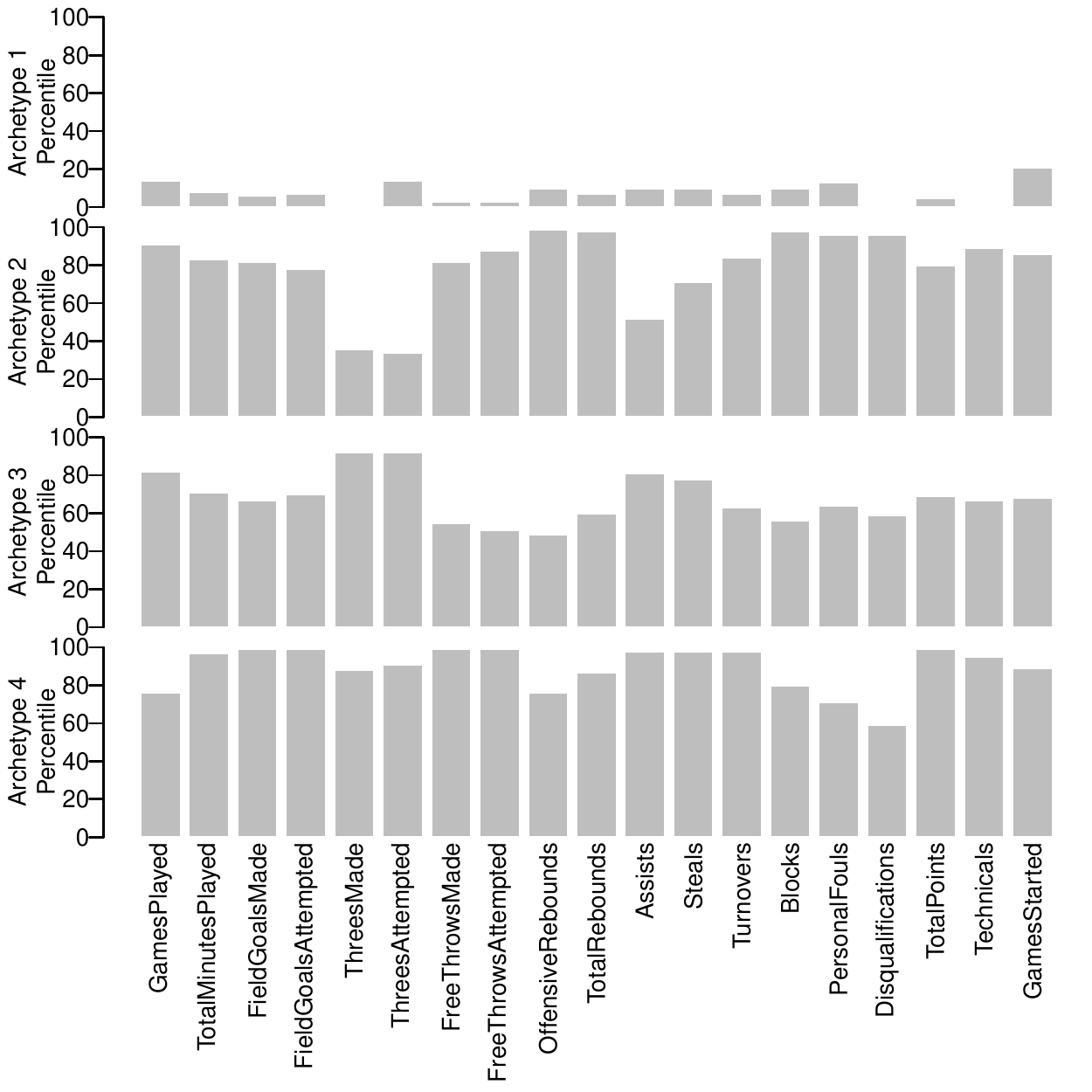}
  \caption{Percentile plot of the four archetypal basketball players
    solution.}
  \label{fig:nba-atypes}
\end{figure}

We determine the archetypal basketball players of the NBA season
2009/2010. \citet{Kubatko+Oliver+Pelton+Rosenbaum@2007} define basic
variables used in what is now the mainstream of basketball
statistics. Following their suggestion we use a data set provided by
\citet{web:dougstats.com} with $19$ statistics of
$441$ players.

Figure~\ref{fig:nba-data} visualizes the data using a parallel
coordinates plot. In comparison to the two-dimensional illustrative
example no structure is easily observable; and there is, for example,
no player which is the maximum over all statistics. We fit $k = 1,
\ldots, 10$ archetypes; Figure~\ref{fig:nba-rss} shows the
corresponding scree plot: the first ``elbow'' is at $k = 4$ ($RSS =
0.04$), the second one at $k = 8$ ($RSS =
0.03$). The additional error reduction between $k = 4$
and $k = 8$ is marginal and we decide on $k = 4$ archetypal basketball
players.

Figure~\ref{fig:nba-atypes} displays the percentile plots (i.e., the
percentile value in an archetype as compared to the data) of the four
archetypal basketball players available in the NBA season
2009/2010. The particular characteristics are:
\begin{description}
  \item [Archetype~1] is the archetypal ``benchwarmer'' with few games
    played and therefore low values in all statistics.

  \item [Archetype~2] is the archetypal rebounder and defensive player
    with high values in the rebounds, blocks and foul-related
    statistics, and low values in the three-pointers.

  \item [Archetype~3] is the archetypal three-point shooter with
    high values in the three-pointer statistics and low values in the
    free throws and rebounds.

  \item [Archetype~4] is the archetypal offensive player with high
    values in all throw-related statistics and low values in
    foul-related statistics.
\end{description}
Archetype~1 represents a type of ``\textit{bad}'' basketball player
while all others represent different types of ``\textit{good}''
players. The four basketball player nearest to one of the four
archetypes are:
\begin{center}
% latex table generated in R 2.13.1 by xtable 1.5-6 package
% Thu Oct 06 13:44:45 2011
\begin{tabular}{rlllrrrr}
  \hline
 & Name & Team & Role & $\alpha_{\cdot1}$ & $\alpha_{\cdot2}$ & $\alpha_{\cdot3}$ & $\alpha_{\cdot4}$ \\
  \hline
Archetype 1 & Dwayne Jones & PHO & C & 1.00 & 0.00 & 0.00 & 0.00 \\
  Archetype 2 & Taj Gibson & CHI & SF & 0.00 & 1.00 & 0.00 & 0.00 \\
  Archetype 3 & Anthony Morrow & GSW & SG & 0.00 & 0.00 & 0.96 & 0.04 \\
  Archetype 4 & Kevin Durant & OKL & SF & 0.00 & 0.00 & 0.00 & 1.00 \\
   \hline
\end{tabular}\end{center}
On this account, Taj Gibson, Anthony Morrow, Kevin Durant can be considered as the best
basketball players of the season 2009/2010 with respect to the
characteristics of their corresponding archetypes. However, note that
in case of Archetype~3 the player is not exactly the archetype. In
order to find all good players, we look at the observations where one
of the three ``\textit{good}'' archetype contributes more than
$0.95$:
\begin{center}
% latex table generated in R 2.13.1 by xtable 1.5-6 package
% Thu Oct 06 13:44:45 2011
\begin{tabular}{llllrrrr}
  \hline
Archetype & Name & Team & Role & $\alpha_{\cdot1}$ & $\alpha_{\cdot2}$ & $\alpha_{\cdot3}$ & $\alpha_{\cdot4}$ \\
  \hline
Archetype 2 & Taj Gibson & CHI & SF & 0.00 & 1.00 & 0.00 & 0.00 \\
   & Andrew Bogut & MIL & C & 0.00 & 1.00 & 0.00 & 0.00 \\
   & Samuel Dalembert & PHI & C & 0.02 & 0.98 & 0.00 & 0.00 \\
   & Jason Thompson & SAC & PF & 0.03 & 0.96 & 0.00 & 0.00 \\
  Archetype 3 & Anthony Morrow & GSW & SG & 0.00 & 0.00 & 0.96 & 0.04 \\
   & Steve Blake & LAC & PG & 0.02 & 0.00 & 0.96 & 0.02 \\
  Archetype 4 & Kevin Durant & OKL & SF & 0.00 & 0.00 & 0.00 & 1.00 \\
   & Lebron James & CLE & SF & 0.00 & 0.00 & 0.00 & 1.00 \\
   & Dwyane Wade & MIA & SG & 0.00 & 0.00 & 0.00 & 1.00 \\
   & Kobe Bryant & LAL & SG & 0.03 & 0.00 & 0.00 & 0.97 \\
   \hline
\end{tabular}\end{center}
The equal coefficients, e.g. for the first two players in case of
Archetype~2, occur due to rounding to two decimal places. The
threshold $0.95$ is arbitrarily defined; this is, in
fact, the only subjective decision one has to make when discussing the
quality of athletes using archetypal analysis.

\subsection{Archetypal soccer players\label{soccer}}

The skill ratings are from the \citet{web:pesstatsdatabase.com} (PSD),
a community-based approach to create a database with accurate
statistics and skill ratings for soccer players (originally for the
video game ``Pro Evolution Soccer'' by Konami). The extracted data set
consists of $25$ skills of $1658$ players
(all positions---Defender, Midfielder, Forward---except Goalkeepers)
from the German Bundesliga, the English Premier League, the Italian
Serie A, and the Spanish La Liga. The skills are rated from 0 to 100
and describe different abilites of the players: physical abilities
like balance, stamania, and top speed; ball skills like dribble, pass,
and shot accuracy and speed; and general skills like attack and
defence performance, technique, aggression, and teamwork. Note that we
assume that the differences are interpretable, i.e., the ratings are
on a ratio scale.

\begin{figure}[t]
  \centering
  \includegraphics{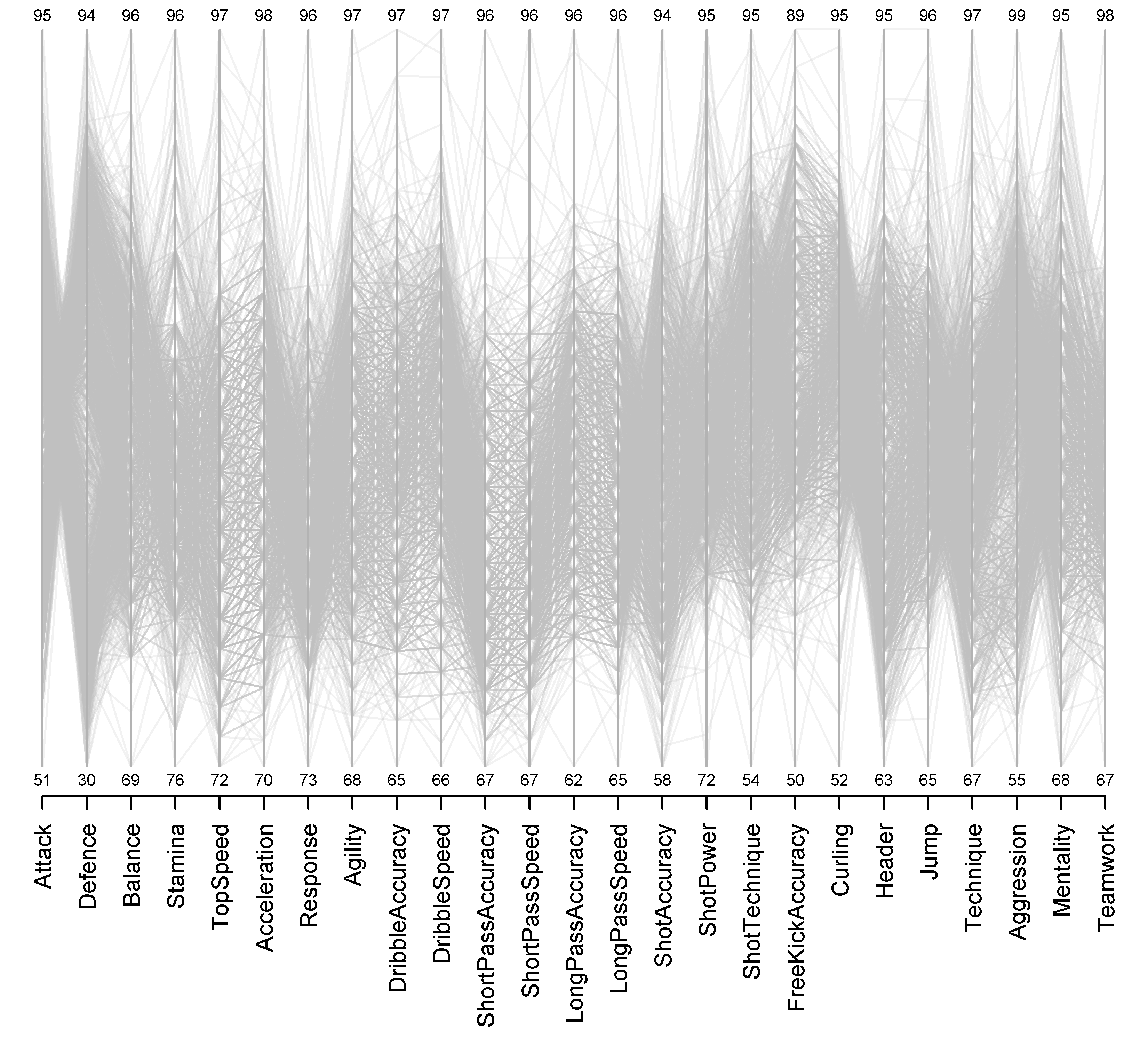}
  \caption{Parallel coordinates of the skill ratings of
    of $1658$ players from the German Bundesliga, the
    English Premier League, the Italian Serie A, and the Spanish La
    Liga.}
  \label{fig:soccer-data}
\end{figure}

Figure~\ref{fig:soccer-data} shows a parallel coordinates plot of the data
set. Most skills range between 50 and 100; this is due to the fact
that PSD describes soccer players of all hierarchy levels of a league
system. Anyway, no real structure is visible in the data, and there
are no players which are the maximum or the minimum over all
skills. We decide to use $k = 4$ archetypal soccer players; see the
online supplement for the decision process (section on computational
details on page~\ref{compdetails}).

\begin{figure}
  \centering
\includegraphics{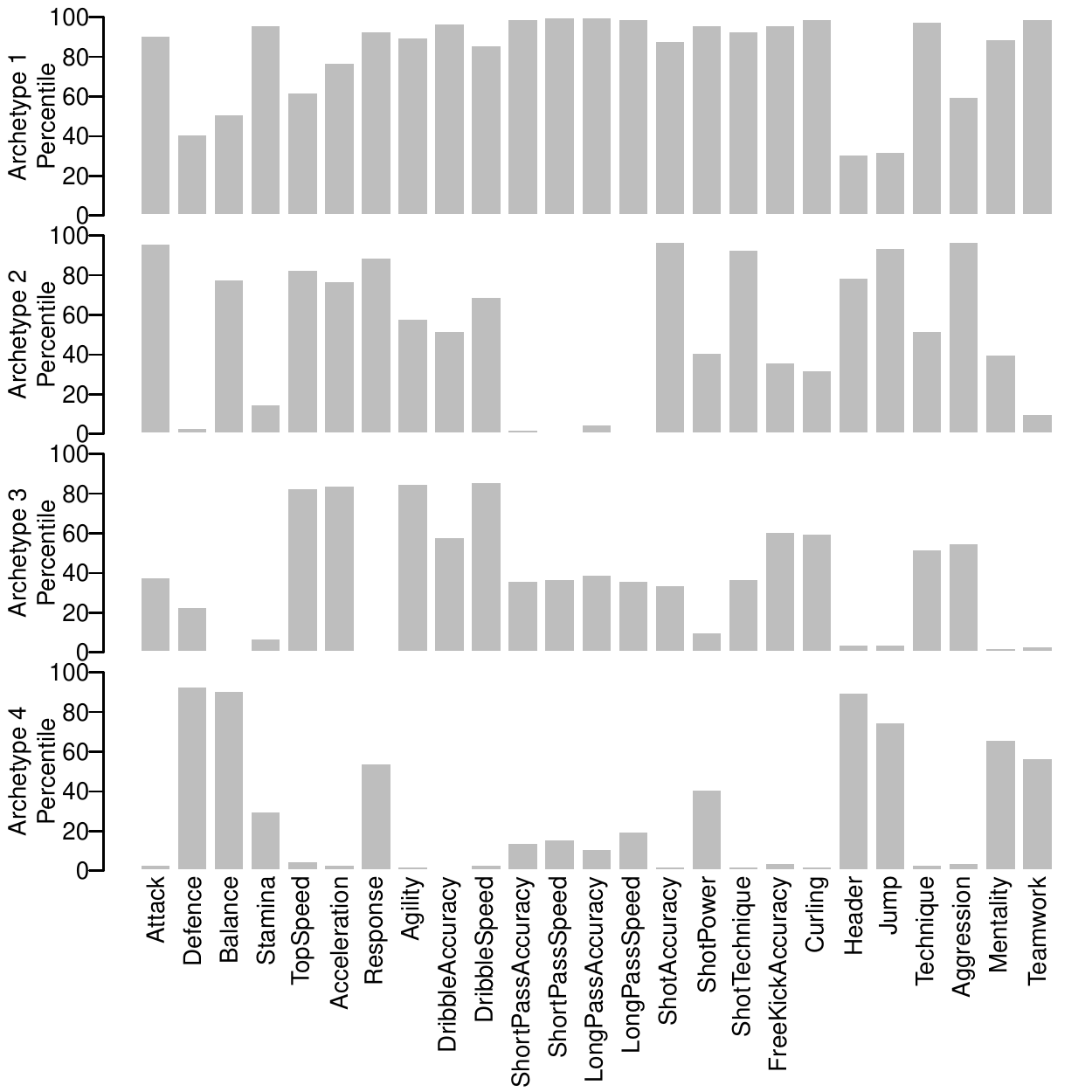}
  \caption{Percentile plot of the four archetypal soccer players
    solution.}
  \label{fig:soccer-atypes}
\end{figure}

\begin{figure}
  \centering
  \includegraphics{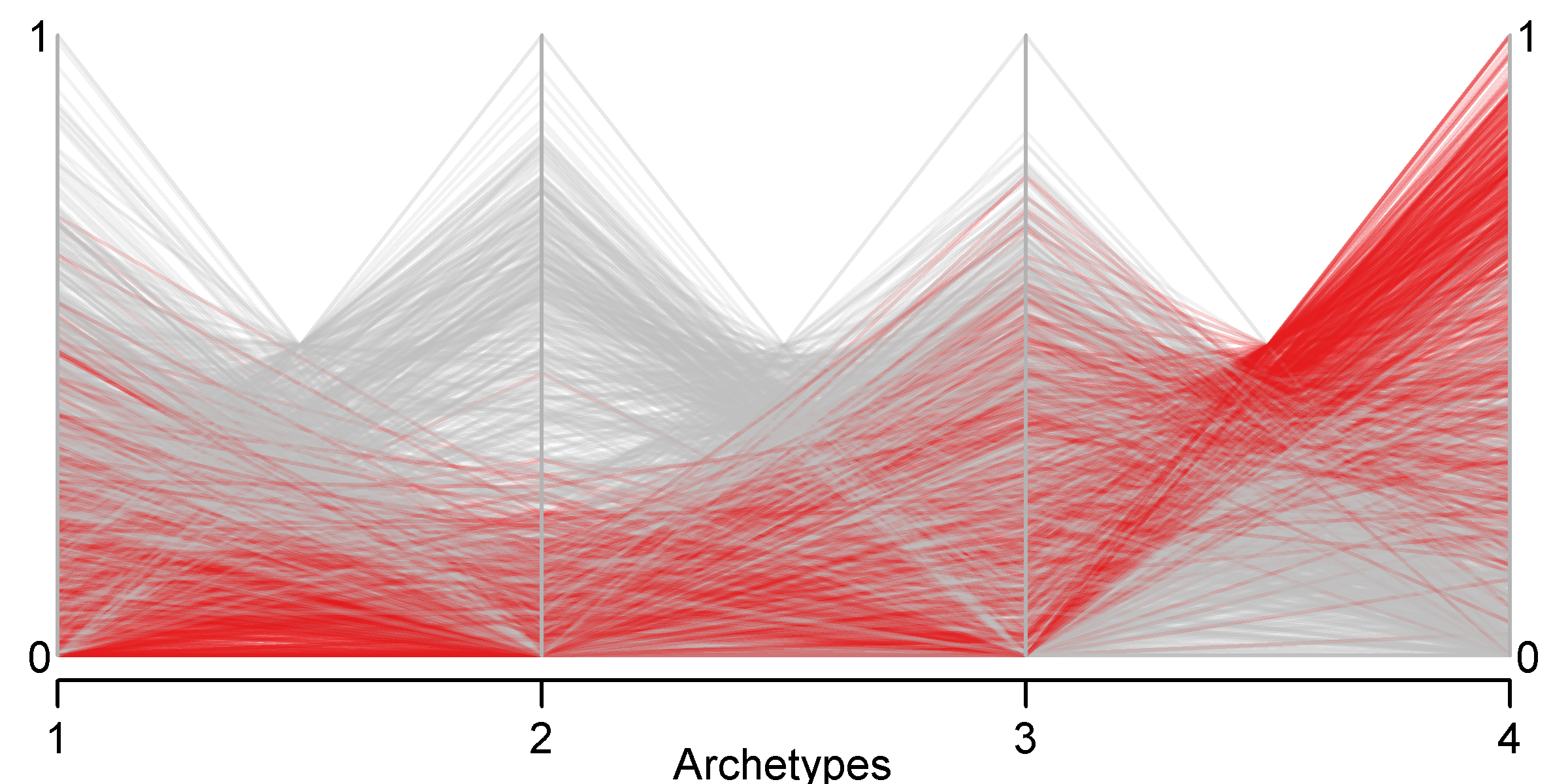}
  \caption{Parallel coordinates plot of $\alpha$ coefficients of the
    four archetypal soccer players solution with highlighted defenders
    (red).}
  \label{fig:soccer-coef}
\end{figure}

Figure~\ref{fig:soccer-atypes} displays the percentile plots of the
four archeypal soccer players. The particular characteristical skills
are:
\begin{description}
  \item [Archetype~1] is the archetypal offensive player with all
    skills high excpect the defense, balance, header, and jump.

  \item [Archetype~2] is the archetypal center forward with high
    skills in attack, shot, acceleration, header and jump, and low
    passing skills.

  \item [Archetype~3] is the archetypal weak soccer player with high
    skills in running, but low skills in most ball related skills.

  \item [Archetype~4] is the archetypal defender with high skills in
    defense, balance, header, and jump.
\end{description}
To verify this interpretation we look at the $\alpha$ coefficients in
combination with the players' position; Figure~\ref{fig:soccer-coef}
exemplarily shows the parallel coordinates plot with the ``Defender''
position highlighted (red). As we can see, nearly all defenders have a
high $\alpha$ coefficient for Archetype~4.

Now, in order to investigate the question of the best soccer player we
have to make a (subjective) definition of ``the best'' in terms of the
four archetypes. For us, the best player is a combination of
Archetype~1 and Archetype~2 with Archetype~1 contributing more than
Archetype~2 (according to the common sense that offensive players are
match-winning). The following soccer players apply to the definition
(orderd according to $\alpha_{\cdot1}$):
\begin{center}
% latex table generated in R 2.13.1 by xtable 1.5-6 package
% Thu Oct 06 13:27:39 2011
\begin{tabular}{llrrrr}
  \hline
Name & Club & $\alpha_{\cdot1}$ & $\alpha_{\cdot2}$ & $\alpha_{\cdot3}$ & $\alpha_{\cdot4}$ \\
  \hline
Wayne Rooney & Manchester United FC & 0.82 & 0.18 & 0.00 & 0.00 \\
  Leo Messi & FC Barcelona & 0.79 & 0.21 & 0.00 & 0.00 \\
  Cristiano Ronaldo & Real Madrid CF & 0.68 & 0.32 & 0.00 & 0.00 \\
  Antonio Di Natale & Udinese Calcio & 0.67 & 0.33 & 0.00 & 0.00 \\
  Carlos Tivez & Manchester City FC & 0.66 & 0.34 & 0.00 & 0.00 \\
  Diego Forlan & FC Internazionale Milano & 0.64 & 0.36 & 0.00 & 0.00 \\
  Dimitar Berbatov & Manchester United FC & 0.60 & 0.40 & 0.00 & 0.00 \\
  Adrian Mutu & AC Cesena & 0.60 & 0.40 & 0.00 & 0.00 \\
  Zlatan Ibrahimovic & AC Milan & 0.54 & 0.46 & 0.00 & 0.00 \\
  Luis Suarez & Liverpool FC & 0.53 & 0.47 & 0.00 & 0.00 \\
  Mladen Petric & Hamburger SV & 0.53 & 0.47 & 0.00 & 0.00 \\
  Xavi Hernandez & FC Barcelona & 0.52 & 0.48 & 0.00 & 0.00 \\
  Didier Drogba & Chelsea FC & 0.52 & 0.48 & 0.00 & 0.00 \\
  Giuseppe Rossi & Villarreal CF & 0.51 & 0.49 & 0.00 & 0.00 \\
   \hline
\end{tabular}\end{center}
Based on our definition and the given skill rating data set, Wayne
Rooney is the best player, followed by Leo Messi and Cristiano
Ronaldo.

%%%%%%%%%%%%%%%%%%%%%%%%%%%%%%%%%%%%%%%%%%%%%%%%%%%%%%%%%%%%%%%%%%%%%%
\section{Conclusion\label{conclusion}}

The present paper applies the statistical method archetypal analysis
to sports data. This enables to compute outstanding---positively
and/or negatively---athletes, i.e., the archetypal
athletes. Statements like ``Dirk Nowitzki is the best
basketball player. No, it's Kevin Durant!'' can be then discussed
completely data-driven and with a well-defined and reproducible amount
of subjectivity. The proposed way is (1)~to estimate the archetypes,
i.e., the archetypal athletes, then (2)~to identify the athletes as
different types of ``\textit{good}'' and ``\textit{bad}'' athletes,
and finally (3)~to set all athletes in relation to the archetypes
(using the $\alpha$ coefficients). The two examples---basketball and
soccer---shows that this is an appropriate approach; the estimated
archetypal athletes definitely are consistent with the general
opinion.

%%%%%%%%%%%%%%%%%%%%%%%%%%%%%%%%%%%%%%%%%%%%%%%%%%%%%%%%%%%%%%%%%%%%%%
\section*{Computational details\label{compdetails}}

All computations and graphics have been done using the statistical
software \textsf{R} 2.13.1 \citep{R}, the \textsf{archetypes}
package \citep{rpkg:archetypes}, and the \textsf{SportsAnalytics}
package \citep{rpkg:SportsAnalytics}. \textsf{R} itself and all
packages used are freely available under the terms of the General
Public License from the Comprehensive R Archive Network at
\url{http://CRAN.R-project.org/.}

Data sets and source codes for replicating our analyses are available
in the \textsf{SportsAnalytics} package. An individual analysis is
executed via  (replace \texttt{***} with \texttt{nba-2d}, \texttt{nba}
and \texttt{soccer}):
\begin{verbatim}
R> demo("archeplayers-***", package = "SportsAnalytics")
\end{verbatim}
The source code file for a demo is accessible via:
\begin{verbatim}
R> edit(file = system.file("demo", "archeplayers-***.R",
+                          package = "SportsAnalytics"))
\end{verbatim}

\bibliographystyle{plainnat}
\bibliography{references}

\bigskip

\textbf{Affiliation:}\\
Manuel J. A. Eugster\\
Institut f{\"u}r Statistik\\
Ludwig-Maximilians-Universtit{\"a}t M{\"u}nchen\\
80530 Munich, Germany\\
E-mail: \url{Manuel.Eugster@stat.uni-muenchen.de}\\
Website: \url{http://www.statistik.lmu.de/~eugster/}

\end{document}